\input{aipcheck}

\documentclass[
    ,final            
  ]
  {aipproc}

\layoutstyle{6x9}

\begin{document}

\title{Charm Input for Determining $\gamma/\phi_3$}

\classification{13.66.Jn, 14.40.Lb, 13.25.Ft}
\keywords      {Hadronic Charmed Meson Decays, CKM Angle $\gamma/\phi_3$}

\author{Peter Zweber (on behalf of the CLEO Collaboration)}{
  address={University of Minnesota, Minneapolis, MN 55455 USA}
}

\begin{abstract}
Overconstraining the CKM unitarity triangle with precision measurements
of its angles continues to test the validity of the Standard Model.
One of these angles, $\gamma/\phi_3$, has been measured by studying 
$B^{\pm} \rightarrow D K^{\pm}$ decays, where 
$D = D^{0}$ or $\overline{D}^{0}$.  
I present recent results of hadronic $D$ decays which will improve 
the sensitivity of $\gamma/\phi_3$ measurements.
\end{abstract}

\maketitle


\section{Introduction}

Measuring the angles of the CKM Unitarity Triangle (UT) is an important 
way to study weak interactions in the Standard Model (SM) and to search 
for Physics Beyond the Standard Model (BSM).  
According to Ref. \cite{CKMfitter}, 
the angles of the UT are $\alpha/\phi_1 = (88.2^{+6.2}_{-4.8})^{\circ}$, 
$\beta/\phi_2 = (21.11^{+0.94}_{-0.92})^{\circ}$, 
and $\gamma/\phi_3 = (70^{+27}_{-29})^{\circ}$.  
The angle $\gamma$ is the most poorly measured UT angle.  
Decreasing the uncertainty on $\gamma$ may show that the UT 
is not closed which would lead to BSM.

The angle $\gamma$ can be determined by measuring the interference between 
$b \rightarrow u$ and $b \rightarrow c$ transitions in 
$B \rightarrow D K$ decays\footnote{Unlabeled particles refer to 
charged particles unless otherwise indicated, 
$D$ = $D^0$ or $\overline{D}^0$, and 
use of charge conjugate modes is implied unless otherwise indicated}, 
with the $D$ decaying to the same final state \cite{GLW}.  
(The decay $B \rightarrow D K$ refers to 
$B^{\pm} \rightarrow D K^{\pm}$, $B^{\pm} \rightarrow D^{\ast} K^{\pm}$, 
$B^{\pm} \rightarrow D K^{\ast\pm}$, and 
$B^{0} \rightarrow D K^{\ast0}$ decays.)  
The Atwood-Dunietz-Soni (ADS) method \cite{2bodyADS} 
suggests to study flavored charm 
decays, e.g., $D \rightarrow K\pi, K\pi\pi\pi,~{\rm and}~K\pi\pi^0$, 
while Giri \emph{et al.} \cite{Giri2003} and 
Bondar and Poluektov \cite{Bondar} suggest to study 
$D \rightarrow K_S^0 \pi^+ \pi^-$ decays 
using a binned model-independent method.

In these proceedings, I present the measurements 
from the CLEO collaboration of the 
strong phase in $D \rightarrow K \pi$ decays, coherence factors for 
$D \rightarrow K \pi \pi \pi$ and $D \rightarrow K \pi \pi^0$ decays, 
and $D^0 \rightarrow K^0 \pi^+ \pi^-$ decays 
using a binned Dalitz plot analysis.  
Measurements of these quantities 
are possible due to the $D^0\overline{D}^0$ pair 
from $\psi(3770)$ decays being produced 
in a quantum correlated state \cite{QC}, 
i.e., the $D^0$ and $\overline{D}^0$ have opposite CP 
since the parent $\psi(3770)$ has C = -1.

\section{Strong Phase in $D \rightarrow K \pi$ Decays}

In $B \rightarrow D(K\pi) K$ decays, 
two of the four final states can have large CP-asymmetries.  
The partial width for a final state with a large CP-asymmetry is 
\begin{equation}
\Gamma(B^- \rightarrow (K^+ \pi^-)_D~K^-) \propto r^2_B + (r^{K\pi}_D)^2 
+ 2~r_B~r^{K\pi}_D~{\rm cos}(\delta_B + 
\delta^{K\pi}_D - \gamma), 
\label{eq:KPi}
\end{equation}
where 
$\langle K^+ \pi^- | D^0 \rangle/\langle K^+ \pi^- | \overline{D}^0 \rangle$ 
= $r^{K\pi}_D~e^{i\delta^{K\pi}_D}$ and $r^{K\pi}_D$ and $\delta^{K\pi}_D$ 
are the magnitude and strong phase difference, respectively, between the 
interfering  $D^0 \rightarrow K^+ \pi^-$ and 
$\overline{D}^0 \rightarrow K^+ \pi^-$ amplitudes.  The variables $r^2_B$ 
and $\delta_B$ are similarly defined for the $B \rightarrow D K$ decay.  
It is possible to constrain $\gamma$ if $\delta^{K\pi}_D$ is known precisely.

Using a 281 pb$^{-1}$ data sample collected at $\psi(3770)$, 
CLEO \cite{CLEODToKPi} measured $\delta^{K\pi}_D$ for the first time 
using single and double tagged yields of $D$ mesons.  
Four CP-even and three CP-odd tag modes were used, along 
with inclusive semileptonic $Xe\nu_e$ decays for flavor-tagging.
These yields were used to determine the mixing parameters 
$x \equiv (M_2 -M_1)/\Gamma$ and $y \equiv (\Gamma_2 -\Gamma_1)/2\Gamma$, 
where $M_{1,2}$ and $\Gamma_{1,2}$ are the masses and widths, respectively, 
of the CP-odd ($D_1$) and CP-even ($D_2$) $D$ meson mass eigenstates and 
$\Gamma \equiv (\Gamma_1 + \Gamma_2)/2$.  
External measurements of $(r^{K\pi}_D)^2$, $x$, $y$, 
$y^{\prime} \equiv y{\rm cos}\delta^{K\pi}_D -  x{\rm sin}\delta^{K\pi}_D $, 
and 
$x^{\prime2} \equiv (y{\rm sin}\delta^{K\pi}_D -  x{\rm cos}\delta^{K\pi}_D)^2$ 
were used to determine $(r^{K\pi}_D)^2$, $x$, $y$, 
$r^{K\pi}_D{\rm cos}\delta^{K\pi}_D$, and 
$r^{K\pi}_D x {\rm sin}\delta^{K\pi}_D$ from a least-squared fit \cite{SunFit}.  
The result is $\delta^{K\pi}_D = (22^{+11}_{-12}(stat)^{+9}_{-11}(syst))^{\circ}$.  
This result will be improved by including more tag modes 
in the analysis of the full 818 pb$^{-1}$ $\psi(3770)$ data sample.

\section{Coherence Factors in $D \rightarrow K \pi \pi \pi$ 
         and $D \rightarrow K \pi \pi^0$ Decays}

The ADS method can be extended to multi-body 
flavor-tagged $D$ decays \cite{MultiBodyADS}, 
which have larger branching fractions.  
Intermediate resonances in multi-body $D$ decays have many contributing 
amplitudes with each point in phase space having its own relative strong phase.  
If particular intermediate resonances are not isolated, then the interference 
term is diluted by a coherence factor, 
e.g., $R_{K3\pi}$ for the decay  $D \rightarrow K \pi \pi \pi$.  
The partial width for a $B \rightarrow D(K\pi\pi\pi) K$ decay 
with a large CP-asymmetry is
\begin{equation}
\Gamma(B^- \rightarrow (K^+ \pi^- \pi^- \pi^+)_D~K^-) \propto 
r^2_B + (r^{K3\pi}_D)^2 
+ 2~r_B~r^{K3\pi}_D~R_{K3\pi}
~{\rm cos}(\delta_B + \delta^{K3\pi}_D - \gamma), 
\label{eq:K3Pi}
\end{equation}
where $R_{K3\pi}$ is confined to the range 0-1.  
If $R_{K3\pi}$ is small, the $D$ decays via 
several significant intermediate decay modes.  
If $R_{K3\pi}$ is large, the $D$ decay 
is dominated by one intermediate decay.  
Analogous parameters exist for other $D$ decays, 
e.g., $D \rightarrow K \pi \pi^0$. 

Using the full 818 pb$^{-1}$ $\psi(3770)$ data sample, 
CLEO \cite{CLEOCohFactor} measured $R_F$ and $\delta^F_D$, 
where $F = K \pi \pi \pi$ and $K \pi \pi^0$, 
using double tagged yields of $D$ mesons.  
The results are 
$R_{K3\pi} = 0.33^{+0.20}_{-0.23}$ and 
$\delta^{K3\pi}_D = (114^{+26}_{-23})^{\circ}$ 
for $D \rightarrow K \pi \pi \pi$ and 
$R_{K\pi\pi^0} = 0.84\pm0.07$ and 
$\delta^{K\pi\pi^0}_D = (227^{+14}_{-17})^{\circ}$ 
for $D \rightarrow K \pi \pi^0$ using external measurements 
of $r^F$ and the mixing parameters $x$ and $y$.  
These coherence factor measurements imply that 
$B \rightarrow D(K\pi\pi^0) K$ decays are sensitive to $\gamma$, 
while $B \rightarrow D(K\pi\pi\pi) K$ decays are sensitive to $r_B$.

\section{Binned Analysis of $D \rightarrow K^0 \pi^+ \pi^-$}

The most precise measurements of the angle $\gamma$ are from 
$B \rightarrow D(K^0_S \pi^+ \pi^-) K$ decays 
since $D \rightarrow K^0_S \pi^+ \pi^-$ decays 
are Cabibbo favored. 
Using 383 million $B\overline{B}$ decays the BaBar collaboration measured 
$\gamma = [76^{+23}_{-24}(stat)\pm5(syst)\pm5(model)]^{\circ}$ \cite{BaBarGamma}, 
and the Belle collaboration determined a preliminary value of 
$\gamma = [76^{+12}_{-13}(stat)\pm4(syst)\pm9(model)]^{\circ}$ \cite{BelleGamma} 
from 657 million $B\overline{B}$ decays.  
The model uncertainties arise from the isobar model analysis 
of flavor-tagged $D \rightarrow K^0_S \pi^+ \pi^-$ decays from 
continuum-produced $D^{\ast\pm} \rightarrow D \pi^{\pm}$ events.

Various authors \cite{Giri2003,Bondar} have proposed to remove 
the model dependence by performing binned analyses of 
the $D \rightarrow K^0_S \pi^+ \pi^-$ Dalitz plot.  
The concept is to divide the Dalitz plot into $2N$ bins, 
ranging from $-N$ to $N$ with bin $N=0$ omitted.  
This segmentation leads to a line of symmetry about 
$M^2(K^0_S \pi^+) = M^2(K^0_S \pi^-)$.  

For $B^{\pm} \rightarrow D(K^0_S \pi^+ \pi^-) K^{\pm}$ decays, 
the number of $B^{\pm}$ events, $N^{\pm}_i$, 
in each bin $i$ of the Dalitz plot is given by 
\begin{equation}
N^{\pm}_i  = 
h_B \left[ K_i + r^2_BK_{-i} + 2r_B \sqrt{K_i K_{-i}}
( c_{i}~{\rm cos}(\delta_B \pm \gamma) 
+ s_{i}~{\rm sin}(\delta_B \pm \gamma)) \right], 
\label{eq:KsPiPi}
\end{equation}
where $K_{i(-i)}$ is the number of flavor-tagged $D$ events 
in bin $i(-i)$, 
and $c_i$ and $s_i$ are the cosine and sine of the phase difference 
$\Delta\delta_D = \delta_D[M^2(K^0_S \pi^+),M^2(K^0_S \pi^-)] 
 - \delta_D[M^2(K^0_S \pi^-),M^2(K^0_S \pi^+)]$.  
Precision measurements of $c_i$ and $s_i$ allow for measurements 
of $\gamma$ with a decreased model uncertainty.  

Using the full 818 pb$^{-1}$ $\psi(3770)$ data sample, CLEO \cite{CLEOGamma} 
has performed a binned Dalitz plot analysis using the method 
suggested by Bondar and Poluektov \cite{Bondar}.  
They measured the strong phase difference using a Dalitz plot 
divided into sixteen bins.  
The bin sizes were determined from the isobar model results 
listed in Ref. \cite{BaBarIsobar}.  
The number of events in bin $i$ of the Dalitz plot in 
$K^0_S \pi^+ \pi^-$ vs. CP tagged events is proportional to $c_i$ 
while the number of events in bins $i$ and $j$ in 
$K^0_S \pi^+ \pi^-$ vs. $K^0_S \pi^+ \pi^-$ tagged events 
are proportional to $c_ic_j$ and $s_is_j$.  
The inclusion of $D \rightarrow K^0_L \pi^+ \pi^-$ decays increased 
the statistics by more than a factor of 2 
but introduced two new parameters. 

The results of $c_i$ and $s_i$ are shown in Table \ref{tab:KsPiPi}.  
CLEO also determined that the model uncertainty in $\gamma$ 
from these results is about $1.7^{\circ}$ based on a toy MC study.  

\begin{table}
\begin{tabular}{crr}
\hline
$i$ & $c_i$ & $s_i$ \\
\hline
0 & $ 0.743\pm0.037\pm0.022\pm0.013$ & $ 0.014\pm0.160\pm0.077\pm0.045$ \\
1 & $ 0.611\pm0.071\pm0.037\pm0.009$ & $ 0.014\pm0.215\pm0.055\pm0.017$ \\
2 & $ 0.059\pm0.063\pm0.031\pm0.057$ & $ 0.609\pm0.190\pm0.076\pm0.037$ \\
3 & $-0.495\pm0.101\pm0.052\pm0.045$ & $ 0.151\pm0.217\pm0.069\pm0.048$ \\
4 & $-0.911\pm0.049\pm0.032\pm0.021$ & $-0.050\pm0.183\pm0.045\pm0.036$ \\
5 & $-0.736\pm0.066\pm0.030\pm0.018$ & $-0.340\pm0.187\pm0.052\pm0.047$ \\
6 & $ 0.157\pm0.074\pm0.042\pm0.051$ & $-0.827\pm0.185\pm0.060\pm0.036$ \\
7 & $ 0.403\pm0.046\pm0.021\pm0.002$ & $-0.409\pm0.158\pm0.050\pm0.002$ \\
\hline
\end{tabular}
\caption{Results for $c_i$ and $s_i$.  The first, second, and third 
uncertainties are statistical, systematic (excluding the effect 
of including $D \rightarrow K^0_L \pi^+ \pi^-$ decays), and the 
systematic from including $D \rightarrow K^0_L \pi^+ \pi^-$ decays.}
\label{tab:KsPiPi}
\end{table}

\section{Conclusion}

CLEO has utilized its sample of $D^0 \overline{D}^0$ pairs produced 
in a quantum-correlated state to measure the strong phase in 
$D \rightarrow K \pi$,  $D \rightarrow K \pi \pi \pi$, and 
$D \rightarrow K \pi \pi^0$ decays, the coherence factor for 
$D \rightarrow K \pi \pi \pi$ and $D \rightarrow K \pi \pi^0$ decays, 
and to perform a binned Dalitz plot analysis of 
$D \rightarrow K^0 \pi^+ \pi^-$ decays.  
All of these measurements can help to constrain 
the CKM angle $\gamma/\phi_3$.  
CLEO is also working to lower the uncertainty on $\delta^{K\pi}_D$ 
and to perform a binned Dalitz plot analysis of 
$D \rightarrow K^0 K^+ K^-$ decays 
using its full 818 pb$^{-1}$ $\psi(3770)$ data sample.  
The BESIII experiment is expected to improve upon these measurements 
once it collects sufficient data at $\psi(3770)$.


\begin{theacknowledgments}
The author wishes to thank the organizers for a simulating conference.  
\end{theacknowledgments}


\begin{thebibliography}{99}

\bibitem{CKMfitter}
J. Charles \emph{et al.} (CKMfitter Group),  
\emph{Eur. Phys. J.} {\bf C41}, 1 (2005), 
http://ckmfitter.in2p3.fr 

\bibitem{GLW}
M. Gronau and D. London, 
\emph{Phys. Lett.} {\bf B253}, 483 (1991);
M. Gronau and D. Wyler, 
\emph{Phys. Lett.} {\bf B265}, 172 (1991).

\bibitem{2bodyADS}
D. Atwood, I. Dunietz, and A. Soni, 
\emph{Phys. Rev. Lett.} {\bf 78}, 3257 (1997); 
D. Atwood, I. Dunietz, and A. Soni, 
\emph{Phys. Rev.} {\bf D63}, 036005 (2001). 

\bibitem{Giri2003}
A. Giri \emph{et al.}, 
\emph{Phys. Rev.} {\bf D68}, 054018 (2003).

\bibitem{Bondar}
A. Bondar and A. Poluektov, 
\emph{Eur. Phys. J.} {\bf C47}, 347 (2006); 
A. Bondar and A. Poluektov, 
\emph{Eur. Phys. J.} {\bf C55}, 51 (2008).

\bibitem{QC}
M. Goldhaber and J.~L. Rosner, 
\emph{Phys. Rev.} {\bf D15}, 1254 (1977);
I.~I.~Y. Bigi and A.~I. Sanda, 
\emph{Phys. Lett.} {\bf B171}, 320 (1986);
Z-Z. Xing, 
\emph{Phys. Rev.} {\bf D55}, 196 (1997);
M. Gronau, Y. Grossman and J. L.Rosner, 
\emph{Phys. Lett.} {\bf B508}, 37 (2001);
D. Atwood and A.~A. Petrov, 
\emph{Phys. Rev.} {\bf D71}, 054032 (2005);
D.~M. Asner and W.~M. Sun, 
\emph{Phys. Rev.} {\bf D73}, 034024 (2006), 
\emph{Phys. Rev.} {\bf D77}, 019901(E) (2008).


\bibitem{CLEODToKPi}
J.~L. Rosner \emph{et al.} (CLEO Collaboration), 
\emph{Phys. Rev. Lett.} {\bf 100}, 221801 (2008); 
D.~M. Asner \emph{et al.} (CLEO Collaboration), 
\emph{Phys. Rev.} {\bf D78}, 012001 (2008).

\bibitem{SunFit}
W.~M. Sun, \emph{Nucl. Instr. Meth.} {\bf A556}, 325 (2006). 


\bibitem{MultiBodyADS}
D. Atwood and A. Soni, 
\emph{Phys. Rev.} {\bf D68}, 033003 (2003). 

\bibitem{CLEOCohFactor}
N. Lowrey \emph{et al.} (CLEO Collaboration), 
subm. to \emph{Phys. Rev. Lett.}, arXiv:0903.4853[hep-ex]. 


\bibitem{BaBarGamma}
B. Aubert \emph{et al.} (BaBar Collaboration), 
\emph{Phys. Rev.} {\bf D78}, 034023 (2008).

\bibitem{BelleGamma}
K. Abe \emph{et al.} (Belle Collaboration), 
arXiv:0803.3375[hep-ex].

\bibitem{CLEOGamma}
R.~A. Briere \emph{et al.} (CLEO Collaboration), 
accepted by \emph{Phys. Rev.} {\bf D}, arXiv:0903.1681[hep-ex].

\bibitem{BaBarIsobar}
B. Aubert \emph{et al.} (BaBar Collaboration), 
\emph{Phys. Rev. Lett.} {\bf 95}, 121802 (2005).

\end{thebibliography}
\end{document}